\documentclass[aps,prl,reprint,superscriptaddress,nobibnotes]{revtex4-2}

\usepackage{amsmath,amssymb,graphicx,color,verbatim,soul,braket}
\usepackage{scalerel,stackengine}
\usepackage{upgreek}
\usepackage{float}
\usepackage{bm}
\usepackage[hidelinks]{hyperref}

\definecolor{mygreen}{rgb}{0,0.5,0}
\definecolor{myblue}{rgb}{0,0,0.75}
\definecolor{mymagenta}{cmyk}{0,1,0,0.12}
\definecolor{mygray}{rgb}{0.5,0.5,0.5}

\begin{document}
	
	\renewcommand{\figurename}{\textbf{Fig.}}
	\title{Quantum-limited generalized measurement for tunnel-coupled condensates}
	
	\author{Maximilian Pr\"ufer}
	\email[Corresponding author: ]{maximilian.pruefer@tuwien.ac.at}
	\affiliation{Vienna Center for Quantum Science and Technology, Atominstitut, TU Wien}
	
	\author{Yuri Minoguchi}
	\affiliation{Vienna Center for Quantum Science and Technology, Atominstitut, TU Wien}
	\affiliation{Institute for Quantum Optics and Quantum Information - IQOQI Vienna,
		Austrian Academy of Sciences}
	\author{Tiantian Zhang}
	\affiliation{Vienna Center for Quantum Science and Technology, Atominstitut, TU Wien}
	\author{Yevhenii Kuriatnikov}
	\affiliation{Vienna Center for Quantum Science and Technology, Atominstitut, TU Wien}
	\author{Venkat Marupaka}
	\affiliation{Vienna Center for Quantum Science and Technology, Atominstitut, TU Wien}
	\author{Jörg Schmiedmayer}
	\affiliation{Vienna Center for Quantum Science and Technology, Atominstitut, TU Wien}

	\date{\today}
	
	\begin{abstract}
		
		The efficient readout of the relevant information is pivotal for quantum simulation experiments. 
		Often only single observables are accessed by performing standard projective measurements. 
		In this work, we implement a generalized measurement scheme based on controlled outcoupling of atoms. 
		This gives us simultaneous access to number imbalance and relative phase in a system of two tunnel-coupled 1D Bose gases, which realize a quantum simulator of the sine-Gordon field theory. We demonstrate that our measurement is quantum limited by accessing number squeezing and show that we can track Josephson oscillation dynamics with the generalized measurements. Finally, we show that the scheme allows the extraction of atoms while maintaining the system's coherent dynamics, which opens up the door to accessing multi-time correlation functions. Our scheme constitutes a step towards accessing quantum properties of the sine-Gordon field theory and, in the future, studying spatially extended systems under continuous monitoring.
	\end{abstract}

	\maketitle
	Accessing the relevant information from quantum many-body systems is crucial for performing reliable quantum simulations \cite{bloch_quantum_2012,eisert_quantum_2015}. Typically, projective measurements of a single observable are performed, and many repetitions of the experiment under comparable conditions provide access to mean values, fluctuations, and even higher-order correlation functions at a single instance in time \cite{schweigler_experimental_2017,rispoli_quantum_2019}.
	{These measurements have two immediate shortcomings: First, correlations between different operators are usually not accessible as they require control over the local measurement basis.
		In this regard, it has been shown that generalized measurements, based on using positive operator valued measures (POVMs) \cite{peres1997quantum}, 
		can lead to new insight in quantum simulation experiments \cite{kunkel_simultaneous_2019,kunkel_detecting_2022,ringbauer2022,FischerICPOVM2022}. 
		Second, due to the destructive nature of the measurement unequal-time correlation functions cannot be accessed.
		However, those two types of correlations are ingredients of utmost importance, amongst others to access entanglement properties \cite{islam_measuring_2015,hauke_measuring_2016,kunkel_detecting_2022,tajik2023verification} or to perform Hamiltonian learning in the quantum regime \cite{ott2024hamiltonian}}.

\begin{figure}[t]
		\centering
		\includegraphics[width=0.99 \linewidth]{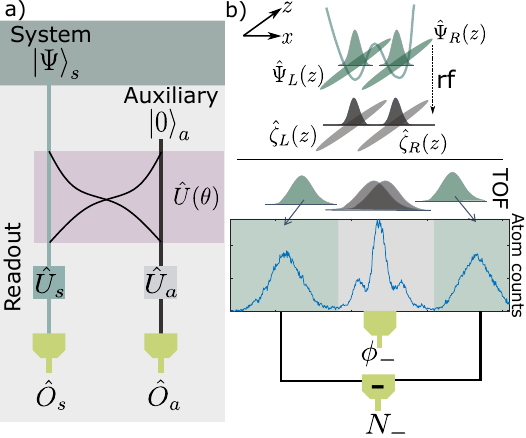}  
		\caption{\textbf{Measurement scheme and implementation.} a) The system is prepared in a state $\ket{\psi}_s$ and coupled via a beamsplitter with an auxiliary system prepared in vacuum $\ket{0}_a$. After the coupling, further manipulations $\hat{U}_{s,a}$ on both systems can be applied, and eventually, both are measured destructively to obtain outcomes associated with operators $\hat{O}_{s,a}$. b) In our case the beamsplitter is realized by radiofrequency radiation coupling two different $m_\text{F}$ states in the $F = 1$ hyperfine manifold of Rubidium. The original system (green) is still trapped while the grey auxiliary system is outcoupled; directly after outcoupling we raise the barrier for the remaining atoms rapidly to apply an outwards momentum. The outcoupled atoms fall under the influence of gravity and form an interference pattern; relative phase, as well as atom number, can be extracted simultaneously from the atom profile.}
		\label{fig:schematic}
	\end{figure}

	The idea of the measurement scheme {which we pursue in this work to address both challenges} is summarized in Fig.\,\ref{fig:schematic} (a). 
	In the system of interest, a quantum state $\ket{\Psi}_s$ is prepared. 
	Before the readout, a beamsplitter operation is performed between the system and an auxiliary system prepared in the vacuum $\vert 0 \rangle_a$. 
	Afterward, unitaries $U_i$ on either system and auxiliary may be applied, and both are then projectively measured{; in principle, $U_s$ can represent an additional time evolution of the system, giving access to unequal-time correlations.}

	{The required beamsplitter operation can be readily implemented in cold atomic systems by coupling of two internal levels.
		In magnetic traps, this coupling transfers atoms to untrapped states, allowing detection without disturbing the remaining atoms. This method was used in early studies of the coherence properties of single component BECs \cite{bloch_measurement_2000,ketterle_coherence_1997, ottl_correlations_2005}. Continuous energy-independent outcoupling can also lead to unconventional cooling in one-dimensional Bose gases \cite{rauer_cooling_2016,grisins2016}. 
		However, so far, the ultimate quantum limits of these measurements and their feedback, and backaction to the system have not been fully utilized and understood.}

	{In this letter we demonstrate a quantum-limited beamsplitter enabling generalized measurements for tunnel-coupled condensates. The {current} scheme allows us to measure the relative phase as well as the relative atom number fluctuations and all their correlations without the need for 
		tomography \cite{gluza2020quantum}. {We explicitly demonstrate that our} beamsplitter implementation does not destroy the quantum state and thus provides a means for performing multiple measurements and the measurement of unequal-time correlation functions in quantum systems in the future.}

	\begin{figure}[t]
		\centering
		\includegraphics[width=1 \linewidth]{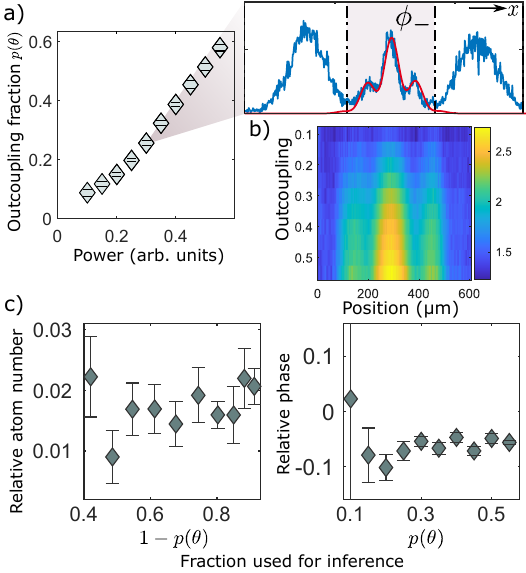}  
		\caption{\textbf{Beamsplitter characterization.} a) Fraction of outcoupled atoms as a function of the outcoupling power. The zoom shows a representative graph of the emerging phase fringe for a power of $0.3$. b) Carpet plot of mean interference fringes as a function of outcoupling strength. c) Inferred relative atom number and relative phase as a function of the fraction of atoms used for inference of each quantity. Over a wide range of outcoupled fractions, we find consistent values for both observables, showing the tunability of the beamsplitter operation.}
		\label{fig:charout}
	\end{figure}

	\textit{Tunable beamsplitter for generalized measurements.---}
	In our experiments, we use elongated Bose-Einstein condensates (BECs) confined by a magnetic trap from an Atomchip [see Fig.~\ref{fig:expseq}(b)] \cite{reichel_atom_2011}. Transversal double well potentials are generated through radiofrequency (RF) dressing \cite{hofferberth_radio-frequency_2006,Lesanovsky2006a}. The beamsplitter for the generalized measurements is implemented using RF radiation, which is typically employed for evaporative cooling. By precisely controlling the frequency, power, and duration of the RF pulse, we achieve a high degree of precision in the outcoupled fraction. For this study, we fix the frequency and duration of the pulse. A short duration results in a pulse with a wide Fourier spectrum, ensuring homogeneous outcoupling along the entire one-dimensional cloud.
	
	Analogous to \cite{kunkel_simultaneous_2019}, we model the outcoupling as a beamsplitter operation on the right and left modes $\hat{\Psi}_{j=R/L}(z)$ of the double well as
	\begin{equation}     \label{Beamsplitter}
		\hat{\Psi}_j(\theta) =  \hat{U}^\dagger(\theta)\hat{\Psi}_j \hat{U}(\theta) =\cos(\theta)\,\hat{\Psi
		}_j + \sin(\theta)\,\hat{\zeta}_{j}.
	\end{equation}
	The angle $\theta$ corresponds to the mixing angle between the condensate and untrapped auxiliary bosonic field operators $\hat{\zeta}_j(z)$. 
	These auxilliary modes are initially unoccupied and thus in the vacuum $\vert 0 \rangle_a = \vert 0\rangle_{a,L} \otimes \vert 0\rangle_{a,R}$, which is annihilated by these operators. 
	
	The effective dynamics of two coupled Bose gases resides in the relative phase of the condensates which is well described in terms of a sine-Gordon field theory \cite{gritsev2007linear}.
	In our measurements, we focus on two global observables:
	Firstly the relative atom number $\hat{N}_- = \int\mathrm{d}z\,\hat{\rho}_-(z)$ of the atoms remaining in the trap, where $\hat{\rho}_-(z) = \hat{\Psi}^\dagger_L(z)\hat{\Psi}_L(z) - \hat{\Psi}^\dagger_R(z)\hat{\Psi}_R(z)$. 
	Secondly the relative phase $\hat{\phi}_-$ of the two condensates which is measured for the outcoupled fraction of the atoms (see Supplementary Information for details).
	The effective dynamics of $\hat{\phi}_-$ and $\hat{N}_-$ corresponds to the dynamics of a damped bosonic Josephson junction \cite{Gati_2007,pigneur2018relaxation}.
	The operators $\hat{N}_-$ and $ \hat{\phi}_-$ are canonically conjugate and the phase space spanned by both will be explored further below.

	{In Fig.~\ref{fig:schematic} (b) we illustrate the measurement scheme; we couple out a fraction $p(\theta)$ of atoms from the trap; we impart an outwards momentum to the remaining atoms (green) before eventually switching off the trapping potential. This separates those atoms during the time-of-flight (TOF) from the interference fringe of the outcoupled atoms (grey) on the fluorescence image \cite{bucker_single-particle-sensitive_2009}.  After TOF, we extract the relative atom number by integrating the atom number profiles in the green regions defined in Fig.~\ref{fig:schematic} (b) which yields $N_-$ for this experimental shot; the relative phase $\phi_-$ is obtained by fitting the interference pattern of the outcoupled atoms in the grey region. }
	
	In Fig.\,\ref{fig:charout} (a), we show the fraction of outcoupled atoms as a function of the pulse power; we find a smooth variation with power. The atomic density from which we deduce the interference fringe is shown exemplarily for $0.3$. 
	We extract the relative atom number and the phase [all fringes shown in in Fig.\,\ref{fig:charout} (b)] and find that above $15\%$  outcoupling expectation values are consistent.
	
	Our measurements show that the mixing angle $\theta$ of our beamsplitter is tunable and robust with respect to the outcoupled fraction. The lower limits of the outcoupled fraction are currently constrained by the signal-to-noise ratio of our imaging setup. Improving this ratio will enable weak probes on smaller outcoupled fractions, allowing for many repeated measurements of the system.

	\begin{figure}[t]
		\centering
		\includegraphics[width=\linewidth]{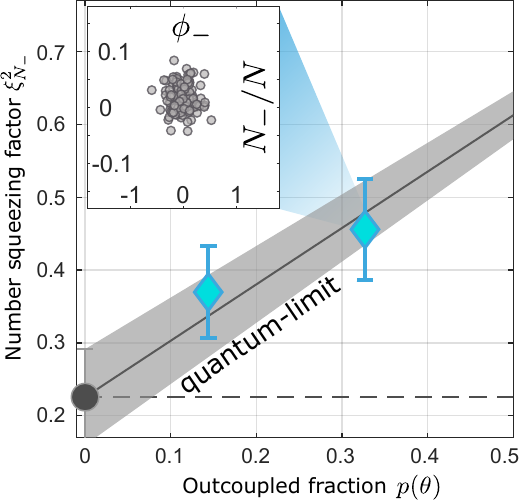}  
		\caption{\textbf{Quantum-limited measurement of number squeezing.} The number squeezing factor $\xi_{N_-}^2(\theta)$ measured after outcoupling for two different fractions; both values below 1 show that we can extract number squeezing with the generalized measurement. The grey data point and the dashed line show the measured number squeezing factor without outcoupling. The grey line plus errorband indicate the expectation for the quantum-limited operation of the readout [according to eq.~\eqref{eq:QLimit}]. 
			The inset shows the measured $(N_-, \phi_-)$ pairs, resembling the phase space distribution of the state.}
		\label{fig:squeezingbackaction}
	\end{figure}
	
	\begin{figure*}[t]
		\centering
		\includegraphics[width=\linewidth]{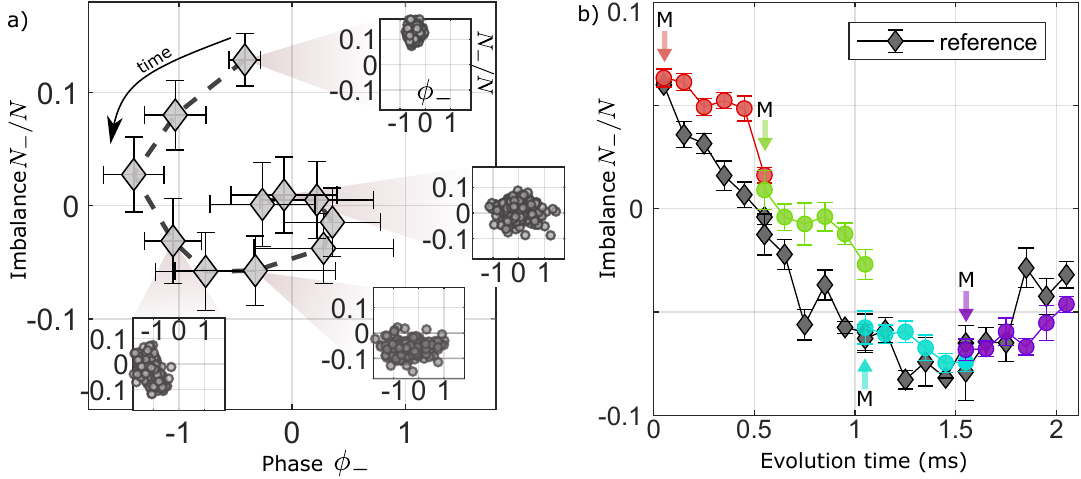}  
		\caption{ \textbf{Josephson oscillations in phase space and post-measurement dynamics.} a) Josephson oscillation of mean values $(\braket{N_-},\braket{\phi_-})$ in classical phase space; direction of time is indicated by the arrow. Insets show the sampled phase space distributions from the acquired experimental realizations, which become accessible through our readout scheme. 
			b) In a different set of measurements, we observe Josephson oscillations in the relative atom number (black diamonds); to probe the measurement backaction on the ensuing dynamics, we couple out a fraction of the atoms at different times (first of colored points (M); $p(\theta) \simeq 0.3-0.4$) and observe the dynamics of the remaining atoms. The subseqent coherent dynamics remain unperturbed except for a change in total atomic density. This suggests that the outcoupling is not destroying the coherence of the quantum state except for adding some noise.}
		\label{fig:Josephson}
	\end{figure*}

	\textit{Quantum-limited operation.---}
	The results from the previous section, where the relative particle number and phase were captured in the very same image, suggest that both observables can be measured simultaneously. 
	However quantum mechanics 
	limits the measurement of two non-commuting observables without incurring some loss of accuracy on both \cite{Braginsky92,peres1997quantum}. 
	Here we demonstrate this quantum limit on the simultaneous readout by focusing on the fluctuation of the particle difference between the left and right wells $\xi^2_{N_-}(\theta) = \Delta N_-^2(\theta)/N(\theta)$ as a function of $\theta$.
	Here $\Delta^2 N_-$ corresponds to the statistical variance of the system before outcoupling. 
	Furthermore we defined the total number of remaining particles $N(\theta) = \langle N_L(\theta)\rangle + \langle N_R(\theta)\rangle $. With this definition $\xi^2_{N_-}(\theta)=1$ corresponds to the standard quantum limit (SQL) \cite{sorensen_many-particle_2001}.
	
	In the Supplementary Information, we show that the relative number fluctuations
	\begin{equation} \label{eq:QLimit}
		\xi^2_{N_-}(\theta) =p(\theta)+  \left( 1-p(\theta) \right)\xi_{N_-}^2  ,
	\end{equation}
	increase linearly with the fraction of outcoupled particles $p(\theta) = \sin^2(\theta)$ and the atom number imbalance is inferred from the remaining $\left( 1-p(\theta) \right)N$ atoms ($N = N(\theta = 0)$). 
	This means that the readout provides access to both quadratures in a single realization allowing for the detection of both number- and spin-squeezing \cite{pezze_quantum_2018}.
	It is interesting to note that our measurement is equivalent to sampling from the Husimi distribution of the quantum state for $p(\theta) = 1/2$ \cite{STENHOLM1992233}.
	
	In our experiment, we routinely produce a type of quantum correlated state known as \textit{number-squeezed} state by splitting a single condensate slowly into a double well potential \cite{Zhang2024}.
	This type of squeezed state is ideally suited for detecting the additional noise introduced by the phase measurement. 
	The variances of the measured observables are estimated by sampling sufficiently many realizations under comparable conditions. 
	In Fig.\,\ref{fig:squeezingbackaction}, we show the detected number squeezing factor with a conventional readout of the relative atom number imbalance as a reference (grey circle and dashed line; error bar is one standard error of the mean). 
	Tuning the power of the outcoupling pulse, we realize two different outcoupling fractions leading to different amounts of introduced additional fluctuations. 
	We observe a detectable noise imprinted by the readout sequence, however, the squeezing is still detectable (turquoise diamonds). 
	We additionally show the sampled points in the phase space spanned by the relative atom number and phase (see Fig.\,\ref{fig:squeezingbackaction} inset).
	
	Our results reveal that our measurement operates right at the quantum limit and can detect the additional {noise stemming from the vacuum fluctuations of the empty port of the beamsplitter;}
	the grey line indicates the theory prediction from eq.\,\eqref{eq:QLimit} (error band is 1 standard error of the mean). 
	{The scheme is rather insensitive to technical fluctuations as they do not affect atom number differences measured for the number squeezing but only introduce fluctuations to $p(\theta)$.}

	\textit{Josephson oscillations.---}
	So far, we have studied static properties for states with fluctuations about zero expectation values; in the following, we study the dynamical evolution of non-zero expectation values.
	To do so, we initiate Josephson oscillations \cite{albiez2005direct,pigneur2018relaxation} and monitor the subsequent time evolution by performing the simultaneous readout.
	In Fig.~\ref{fig:Josephson} (a) we show that we are able to resolve Josephson oscillations, that is we can detect meaningful and consistent non-zero mean values of number and phase.

	Additionally, our new measurement scheme enables sampling from the phase space distribution. The sampled distributions, shown as insets with single samples as grey circles in Fig.~\ref{fig:Josephson} (a), reveal notable deformations, deviating from round, coherent-state-like shapes. Compared to the initial state distribution, we deduce that the shapes result from quantum dynamics and inhomogeneous broadening due to initial state fluctuations, not measurement noise. Our measurements open new avenues for studying dynamical evolution. In the future, it would be interesting to examine the impact of the quantum nature of the local observables in our system, contrasting with the global observables discussed here \cite{pigneur2018relaxation,mennemann_relaxation_2021}.

	\textit{Post-measurement dynamics.---}
	In the following, we want to discuss the possibility of measuring only the outcoupled atoms while evolving the trapped atoms. 
	Our scheme, which combines a long TOF with fluorescence imaging, in principle, allows us to detect the outcoupled atoms without further disturbing the atoms remaining in the trap.
	
	We explicitly tested that the outcoupling does not destroy the subsequent dynamical evolution. 
	For this, we perform a Josephson oscillation experiment; in Fig.~\ref{fig:Josephson}~(b) we show the time evolution of the number imbalance in the unperturbed case as a reference (black data points). 
	At different times $t$ during the evolution, we apply an outcoupling pulse ($p(\theta) \simeq 0.3-0.4$) and observe the ensuing dynamics of the remaining atoms. 
	Averaging over all measurement results of the outcoupled atoms, we observe a dynamical evolution that is slowed down by the change in atomic density (due to the outcoupling; the according plasma frequency scales $\propto \sqrt{N}$) but seems otherwise unperturbed by the measurement protocol.
	
	The ability to resolve and condition on single measurement outcomes will open an entirely new field of inquiry in cold atomic systems \cite{Skinner19,Li18,Bao20,Jian20,Potter2022,Fisher23}. 
	On the quantum level, 
	a measurement on the auxiliary system will also be felt by the atoms that remain trapped, a disruptive event known as \textit{measurement backaction} \cite{Braginsky92}. 
	For the case where at time $t$ the relative phase of the outcoupled modes was found to be $\phi_-$ the auxiliary state collapses to $\vert \phi_- \rangle_a$. 
	Here we probe post-measurement evolution by evolving the state conditioned on the measurement outcome for the duration $\tau$ yielding
	$\vert \Psi(t+\tau\vert t,\phi_-\rangle_s \sim e^{-iH\tau}E_{\phi_-}\vert \Psi(t)\rangle_s$. 
	We defined the operator $E_{\phi_-}$ (a POVM) acting on the system state $\vert \Psi(t)\rangle_s$ encoding the measurement backaction also known as \textit{Kraus} operator \cite{Wilde_2013} as well as the Hamiltonian $H$ which generates the unitary evolution (see Supplementary Information for details).
	The effect of this operator on the subsequent unitary dynamics is crucial when inferring unequal time correlation functions to explore the physics of many-body nonequilibrium quantum dynamics. Furthermore, performing multiple outcoupling pulses and adding unitary time evolution for the remaining atoms in between opens the possibility of studying measurement-induced dynamics and criticality \cite{Li18,Skinner19,Bao20,Jian20,Potter2022,Fisher23,Alberton21}. 
	We are already able to detect the outcoupled atoms while keeping the remaining atoms trapped; magnetic field gradients destroy the interference pattern as atoms still experience a second-order Zeeman effect. 
	This prevents us currently from extracting the relative phase information which so far impedes conditioning the dynamics on the measurement outcome.

	\textit{Outlook.---}
	We have presented a scheme for detecting both quadratures of a bosonic Josephson junction by using controlled outcoupling from magnetic traps. We envision that our scheme will give access to unequal-time correlation functions, which are pivotal for studying thermalization. Further, the possibility of inferring higher-order correlations between the two quadratures will allow the detection of entanglement in non-quadratic systems \cite{gaerttner2023Qentangelment,tajik2023verification,haas2024area} as well as Hamiltonian learning in the quantum regime \cite{ott2024hamiltonian}. Using local outcoupling with artificial magnetic fields \cite{lannig_collisions_2020}, local measurements can be performed, and the spreading of perturbation in the 1D system can be investigated. 
	{Furthermore}, our scheme opens the door to studying the transition from closed to open quantum systems combined with continuous measurements interesting, for example, for quantum sensing \cite{ilias2022contsensing} and measurement induced dynamics and criticality \cite{Li18,Skinner19,Fisher23,Potter2022}.
	
	\begin{acknowledgments}
		\textit{Acknowledgments.---} We thank Igor Mazets, Philipp Kunkel, and Martin Gärttner for discussions. This work is supported by the European Union's Horizon Europe research and innovation program the ERC-AdG {\em Emergence in Quantum Physics} (EmQ) under Grant Agreement No. 101097858 and the DFG/FWF CRC 1225 'ISOQUANT', (Austrian Science Fund (FWF) I~4863). M.P. has received funding from the European Union’s Horizon 2020 research and innovation program under the Marie Skłodowska-Curie grant agreement No.~101032523 and from Austrian Science Fund (FWF): ESP~396 (QuOntM). Y.M. acknowledges funding from the European Research Council (Consolidator grant ‘Cocoquest’ 101043705) and Grant No.~62179 of the John Templeton Foundation.

		~
		
		~
		MP and JS conceived the measurement scheme. MP and TZ took the experimental data with the help of YK. MP and TZ analyzed the experimental data. MP and YM developed the theoretical model. All authors contributed to the discussion of the results. MP and YM wrote the manuscript with input from all authors.
	\end{acknowledgments}

	\appendix
	\section*{Supplementary Information}
	\setcounter{figure}{0}
	\renewcommand{\figurename}{\textbf{Supplementary Information Fig.}}
	\textit{Experimental Sequence.---}
	We initially prepare a single condensate by using evaporative cooling and then split the condensate by slowly ramping up the RF-dressing amplitude [see Supplementary Information Fig.~1]. This allows us to prepare different initial states. To initiate Josephson oscillations \cite{albiez2005direct,pigneur2018relaxation}, we first split from a single well into a tilted double well; this leads to a state with a non-zero mean number imbalance. After the splitting ramp, we equalize the energy of the two wells to observe the oscillations in a balanced double well. To prepare a number squeezed state, we perform a slow ramp to a balanced, slightly coupled double well \cite{Zhang2024}.
	
	\begin{figure}
		\centering
		\includegraphics[width=0.99 \linewidth]{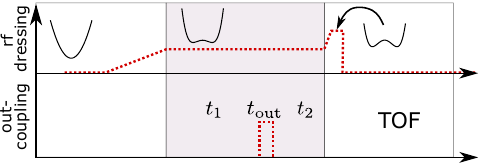} \caption{\textbf{Experimental sequence.} Schematic of experimental timings of RF dressing for producing the double well and the kick, the chip trap, and offset field and the RF for the outcoupling pulse. For data in Fig.~\ref{fig:charout} and~\ref{fig:squeezingbackaction} $t_1=t_2=0$ and $t_\text{out} = 0.025\,$ms; for Fig.~\ref{fig:Josephson}a $t_1$ is varied, $t_2=0$ and $t_\text{out} = 0.025\,$ms; for Fig.~\ref{fig:Josephson}b $t_1$ and $t_2$ are varied and $t_\text{out} = 0.025\,$ms. The total time for the kicking routine is 0.1\,ms.}
		\label{fig:expseq}
	\end{figure}
	
	\textit{Variable Beamsplitter.---} \label{app:BS}
	We apply an intense pulse of RF magnetic fields which couples the trapped dressed state and the untrapped dressed state (see Fig.\,\ref{fig:schematic} (b)). 
	We approximate the outcoupling dynamics by $U(\theta) = \exp(-i\theta H_{\mathrm{out}} )$ where we defined the Hamiltonian $H_{\mathrm{out}} = \sum_{j=L,R}
	\int\mathrm{d}z\,(
	\hat{\zeta}^\dagger_j(z)\hat{\Psi}_j(z) + \mathrm{h.c.}
	)$.
	Here $\hat{\Psi}_j(z)$ ($\hat{\zeta}_j(z)$) correspond to the field operators of the bosonic atoms in the trapped (untrapped) dressed state. 
	The angle of the beamsplitter is then $\theta = \Omega_{\mathrm{rf}} t_{\rm out}$ where $\Omega_{\mathrm{rf}}$ corresponds to the power of the RF pulse and $t_{\rm out}$ to its duration. 
	This outcouples an adjustable amount of atoms that fall under the action of gravity which is described by $\hat{U}_2$ in Fig.~\ref{fig:schematic} (a) of the main paper. 
	
	For accessing the number imbalance
	we rapidly increase the RF dressing amplitude to raise the central barrier of the double well [see Supplementary Information Fig.~1]; this imprints an outwards momentum onto the remaining atoms (see $\hat{U}_1$ in Fig.~\ref{fig:schematic} (a) of main paper) before we switch off all trapping potentials.
	While falling the atomic clouds expand and the outcoupled atoms overlap and form an interference pattern; the kicked, remaining atoms separate such that the interference fringe and the two kicked clouds can be imaged separately on the camera [see Fig.\,\ref{fig:schematic} (c)]. 
	The lower panel shows the integrated atomic densities along the original longitudinal extension of the cloud. Our methods thus not only provide information about global quantities but, in principle, also give spatial resolution along the elongated direction of the Bosonic-Josephson junction.

	\textit{Number and Phase Measurement.---} \label{app:thy_basics}
	The effective theory of one inhomogeneous Bose condensate is developed in \cite{mora_extension_2003}.
	The essence of these treatments relevant for this work is that the dynamics of the excitation on the top of the condensate are most conveniently parametrized by the decomposition of the field operator in terms of density and phase operators, $\hat{\Psi}_j(z) = e^{i\hat{\Phi}_j(z)}\sqrt{\hat{\rho}_{j}(z)}$. 
	The phase and the density operator are canonically conjugate $[\hat{\Phi}_i(z), \hat{\rho}_j(z^\prime)] = i \delta_{ij}\delta(z-z^\prime)$.
	The density operator is readily expressed in terms of the field operators $\hat{\rho}_j(z) = \hat{\Psi}^\dagger_j(z) \hat{\Psi}_j(z)$. 
	In the presence of a quasi-condensate $\hat{\rho}_j(z) = \rho_{0,j}(z) + \delta \hat{\rho}_j(z)$ the density fluctuations relative to the condensate $\rho_{0,j}(z)$ are considered.
	In \cite{gritsev2007linear} the dynamics of two coupled condensates is described by an effective sine-Gordon field theory in $z$-direction with the dynamical degrees of freedom being the relative density (phase) fluctuations $\delta \hat{\rho}_-(z)$ ($\hat{\Phi}_-(z) = \hat{\Phi}_L(z) - \hat{\Phi}_R(z)$) respectively. 
	In our Atomchip experiment, we are able to access both relative phase and density.
	In the interference measurement (after some free expansion in the $(x,z)$-plane) we probe the two overlapping condensates via fluorescence imaging of the total particle density $\hat{\rho}_{\rm tot}(x,z) = (\hat{\Psi}_L^\dagger(x,z) + \hat{\Psi}_R^\dagger(x,z) ) (\hat{\Psi}_L(x,z) + \hat{\Psi}_R(x,z) )$. 
	In the limit of small density fluctuations we obtain $ \hat{\rho}_{\rm tot}(x,z) \simeq \hat{\rho}_L(x,z) + \hat{\rho}_R(x,z) + 2\rho_{0}(x,z)\cos(k_{\rm TOF} x + \hat{\Phi}_-(z))$ where $k_{\rm TOF}$ governs the fringe pattern in $x$-direction after expansion for a given TOF.
	Furthermore we assumed that the condensate is symmetric $\rho_{0} = \rho_{0,j}$ for $j=L,R$.
	
	In this experiment, we consider global observables by marginalizing over the $z$-direction $\hat{\rho}_{\rm tot}(x) = \int \mathrm{d}z\, \rho_{\rm tot}(x,z)$.
	From this we find 
	$\hat{\rho}_{\rm tot}(x) = \hat{\rho}_L(x) + \hat{\rho}_R(x) + 2\rho_0(x) \cos(k_{\rm TOF}x + \hat{\phi}_-)$ and it is possible to identify the (global) relative phase $\phi_-$ from a single measurement run as shown in the red line in Fig.~\ref{fig:charout}~(b).
	Separating the condensates by an outward kick [modeled by $\hat{U}_1$ in Fig.~\ref{fig:schematic}~(a)] makes the interference term vanish to a good approximation and we obtain $\hat{\rho}_{\mathrm{tot}}(x) \simeq \hat{\rho}_L(x) + \hat{\rho}_R(x)$.
	The fluorescence image for a single shot yields the outer fringes (green background) in Fig.~\ref{fig:charout}~(a) and the relative number fluctuation for single realization is obtained from the difference $N_- \simeq  \int \mathrm{d}x\,(\delta\rho_L(x) - \delta\rho_R(x)) $, since left and right condensate densities cancel.

	\textit{Quantum Limit for Simultaneous Detection.---}
	The relative number of particles after coupling out a variable amount of atoms for the phase measurement is described by $\hat{N}_-(\theta) = U^\dagger(\theta)\hat{N}_-U(\theta)$.
	Note that the relative number of particles is $\hat{N}_- = \int\mathrm{d}z\,(\hat{\Psi}_L^\dagger(z) \hat{\Psi}_L(z) -  \hat{\Psi}_R^\dagger(z) \hat{\Psi}_R(z))$.
	Using the action of the beamsplitter on the field operator from Eq.~\eqref{Beamsplitter} we obtain
	\begin{align} \label{eq:Nminus}
		\hat{N}_-(\theta) = & \cos^2(\theta) \hat{N}_- + \sin^2(\theta) \hat{N}_{\zeta,-} \\
		& + \sin(\theta)\cos(\theta) \sum_{j=L,R}\int\mathrm{d}z\, (\hat{\Psi}^\dagger_j(z)\hat{\zeta}_j(z) +\mathrm{h.c.}) \nonumber,
	\end{align}
	where we defined the relative number of the outcoupled field $\hat{N}_{\zeta,-} = \int\mathrm{d}z\,(\hat{\zeta}_L^\dagger(z) \hat{\zeta}_L(z)- \hat{\zeta}_R^\dagger(z) \hat{\zeta}_R(z))$.
	The expectation value is taken with respect to the right state before outcoupling, $\vert \Psi(t)\rangle_s \otimes \vert 0 \rangle_a$.
	Using the annihilation property of $\hat{\zeta}_j(z)\vert 0 \rangle_a = 0$ we obtain $\langle \hat{N}_-(\theta)\rangle = 0$.
	In order to obtain the variance $\Delta^2 N_-(\theta) = \langle N_-^2(\theta)\rangle $ we square the operator in Eq.~\eqref{eq:Nminus} and using
	\begin{align}
		\int\mathrm{d}z^\prime \hat{\Psi}_j(z)\Psi_j(z^\prime) \langle 0 \vert \hat{\zeta}_j(z)\hat{\zeta}_j^\dagger (z^\prime)\vert 0 \rangle_a =\hat{\Psi}^\dagger_j(z)\hat{\Psi}_j(z),
	\end{align}
	which follows from the bosonic commutation relation $[\hat{\zeta}_i(z),\hat{\zeta}_j(z^\prime)] = \delta_{ij}\delta(z-z^\prime)$ of the outcoupled auxiliary modes.
	Finally computing $N(\theta)$ in an analogous way we obtain the final result for $\xi_{N_-}^2(\theta)$.

	\textit{Measurement Backaction and Kraus Operators.---}
	For deriving the Kraus operator, we assume for simplicity that the state of system and auxiliary system at time $t$, right before outcoupling, is $\vert \Psi(t)\rangle_s \otimes \vert 0\rangle_a$.
	As discussed above the outcoupling is modeled by $\hat{U}(\theta)$.
	Given the relative phase $\phi_-$ was measured, the state of the auxiliary system collapses to $\vert \phi_-\rangle_a$ right afterwards.
	The backaction of this collapse acting on $\vert \Psi(t)\rangle_s$ is obtained by considering
	$_a\langle \phi_- \vert U(\theta)\vert \Psi(t)\rangle_s \vert 0 \rangle_a = \hat{E}_{\phi_-}\vert \Psi(t)\rangle_s$.
	Here we defined the (Kraus) operator $E_{\phi_-} = \langle \phi_-\vert \hat{U}(\theta)\vert 0 \rangle_a$ acting on the system wave function $\vert \Psi(t)\rangle_s$.
	Taking into account the correct normalization of the state (which we omitted in the main text) we obtain $\vert \Psi(t+\tau \vert \phi_-,t)\rangle_s = e^{-i\hat{H} \tau} \hat{E}_{\phi_-}\vert \Psi(t)\rangle_s/(\langle \Psi(t)\vert \hat{E}^\dagger_{\phi_-}\hat{E}_{\phi_-}\vert \Psi(t)\rangle_s)^{1/2} $.
	Here we included an additional unitary evolution generated by the Hamiltonian $H$ for duration $\tau$ after the measurement.

\end{document}